\documentclass[aps,pra,preprint,groupedaddress,showpacs]{revtex4}
\usepackage{amsfonts}
\usepackage{graphicx}

\begin{document}

%\preprint{}

\title{Consistency of the adiabatic theorem and perturbation theory}


\author{Marco Frasca}
\email[]{marcofrasca@mclink.it}
%\thanks{}
\affiliation{Via Erasmo Gattamelata, 3 \\ 
00176 Roma (Italy)}


\date{\today}

\begin{abstract}
We present an analysis of the adiabatic approximation to understand when it applies, in view of the recent criticisms and studies for the validity of the adiabatic theorem. We point out that this approximation is just the leading order of a perturbation series, that holds in a regime of a perturbation going to infinity, and so the conditions for its validity can be only obtained going to higher orders in the expansion and removing secular terms, that is terms that runs to infinity as the time increases. In this way, it is always possible to get the exact criteria for the approximation to hold.
\end{abstract}

%\keywords{}

\maketitle


\section{Introduction}

The formulation of the adiabatic theorem is as old as quantum mechanics itself \cite{born}. A sound mathematical formulation was then provided by Kato \cite{kato} with the introduction of the idea of a gap condition and finally, Berry \cite{berry} showed as phases cannot be disposed away so easily. Essentially, it can be stated in a similar way as also applies in mechanics and thermodynamics: A system keeps on staying into its initial state provided its evolution in time is slow enough. To clearly define the terms of this definition one has to rely on the specific framework the system is considered. So, for quantum mechanics, we need to specify a Hamiltonian $H$ that evolves in time. Once we are able to solve the corresponding eigenvalue problem, we get eigenstates varying in time and if we start from one of such eigenstate at the initial time, a slow variation of the Hamiltonian, properly stated through a time scale $T$ large enough, maintains the system in this eigenstate properly evolved in time and multiplied by some phases. Slowness condition can be properly stated and this condition should grant the validity of the given conclusion.

This idea has been so successful that has found a lot of applications but it appears at the foundations of quantum computation \cite{farhi,steff,rola} providing a promising way to implement it. So, it appeared somehow as a surprise the fact that its condition of validity was not sufficient \cite{marz,tong}. Marzlin and Sanders \cite{marz} were able to display an example where the condition for the adiabatic theorem to hold just fails opening up the way to a hot debate that is well alive yet\cite{tong1,mack1,vert,tong2,guo,mack2,wei,zhao,zhao1,comp,amin,nori,wiebe} and we can safely say that this matter is not settled yet.

The aim is to paper is to clarify this question based on a couple of papers appeared in the nineties \cite{most,fra1}. The reason of the failure of the condition of validity for the adiabatic approximation can be traced back on the very nature of the approximation itself. This appears to be just the leading order of a series expansion of the solution of the Schr\"odinger equation. This series, arising as a solution of a differential equation, displays exactly the same defects of a more known small perturbation series. These are a typical expansion parameter and secular terms. A typical expansion parameter always appears in a perturbation series due to the physical nature of the problem at hand. This is fundamental to give a mathematical meaning to the series itself. Secular terms represent a major failure in a series expansion as they just increase without bounds as the independent variable, time in our case, is taken increasingly large. So, before to give a meaning to an expansion as is the adiabatic approximation one has to dispose of such terms. Techniques exist to do this but we do not discuss them here. Rather, we will show how, going to the next-to-leading and next-to-next-to-leading orders, we are able to recover the proper validity conditions for a given adiabatic expansion.

% Added after referee's comments
As we will show, in agreement with what is expected in Ref. \cite{comp}, secular terms are present in a perturbation series independently on resonances being there or not. These kind of unbounded contributions can appear also in the more general case as we will show. This means that our understanding of the failure of the criteria for the adiabatic approximation to hold is quite general.

The paper is so structured. In sec.\ref{sec1} we expose the troubles with the adiabatic approximation as stated in literature. In sec.\ref{sec2} we exploit the deep connections between the adiabatic approximation and perturbation theory and how this approximation provides a dual expansion to the well-known Dyson series. In sec.\ref{sec3} we show how one can dispose of some examples presented in literature, proving the failure of the standard validity condition in the adiabatic approximation, and obtaining the proper validity conditions. Finally, in sec.\ref{sec4} conclusions are given.

\section{\label{sec1}Failure of the adiabatic approximation}

As it is well-known from textbooks \cite{mess}, the adiabatic approximation in quantum mechanics shows how a system keeps on staying on the initial state, provided the parameters of the Hamiltonian are varied slowly enough. A proper mathematical definition of slow evolution should be given for this statement to be considered as a theorem and to have a quantitative understanding of the limitations of this kind of approximation.

The main result can be stated in the following way. Given a quantum system with a Hamiltonian $H(t)$ and letting it varying on a finite interval of time $t\in [0,T]$, the Schr\"odinger equation admits the following approximate solution for the time evolution operator 
% 16 November 2011 Asked by the referee
\cite{most,fra1}
\begin{equation}
\label{eq:ua}
   U(t,0)\approx\sum_n e^{i\gamma_n(t)}e^{-\frac{i}{\hbar}\int_0^t dt'E_n(t')}|n;t\rangle\langle n;0|=U_0(t,0)
\end{equation}
provided we are able to solve the eigenvalue problem $H(t)|n;t\rangle=E_n(t)|n,t\rangle$ and being $\dot\gamma_n(t)=\langle n;t|i\partial_t|n;t\rangle$ the geometrical phase. It is common practice to absorb this latter phase into the eigenstates but, as we will show below, this practice should be dismissed. This holds if the following condition holds \cite{born,kato}
% Corrected as asked by the referee 30 January 2012
\begin{equation}
\label{eq:cond}
    \sum_{n\ne m}\frac{\hbar}{|E_n(t)-E_m(t)|}\left|\frac{\langle n;t|\dot H(t)|m;t\rangle}{E_n(t)-E_m(t)}\right|\ll 1,
\end{equation}
%being $n\ne m$ and 
%
provided a gap exists between the eigenvalues $E_n(t)$ and $E_m(t)$. Recently, this condition has been proved to be neither sufficient nor necessary marking evidence for the failure of the adiabatic approximation even if it seems to hold \cite{marz,tong}. We will discuss several examples of this in sec.\ref{sec3}, taken from the literature, and we will also show how to get the right adiabaticity condition.

\section{\label{sec2}Adiabatic approximation and perturbation theory}

%\subsection{\label{ss2}Theoretical analysis}

Adiabatic approximation is generally stated for slowly varying Hamiltonians\cite{mess}. Typically, one has to solve a time-dependent problem like
\begin{equation}
   H(t)|\psi\rangle=i\hbar\partial_t|\psi\rangle.
\end{equation}
Now, varying $t$, one has a continuous set of Hamiltonians and if the system was in the eigenstate $|n;0\rangle$ of $H(0)$, the adiabatic theorem states that the system stays in the same eigenstate $|n;t\rangle$ of $H(t)$ at later times, except for a phase factor. This can be also seen by introducing a time scale $T$ so that our problem to solve can be stated as
% Modified as required by referee 30 January 2012
\begin{equation}
   H(t,T)|\psi\rangle=i\hbar\partial_t|\psi\rangle.
\end{equation}
If $T$ is taken large enough, on smaller time scales the system is adiabatic. One can apply a change of time variable as $\tau=t/T$ and write down the above equation as
\begin{equation}
\label{eq:T}
   TH(\tau,T)|\psi\rangle=i\hbar\partial_\tau|\psi\rangle.
\end{equation}
and the solution of this equation for $T\rightarrow\infty$, at the leading order, is exactly the statement of the adiabatic theorem. This can also be said by approximating the unitary evolution of the original problem as in eq.(\ref{eq:ua}).
%\begin{equation}
%   U(t,0)\approx U_0(t,0)=\sum_n e^{i\gamma_n(t)-\frac{i}{\hbar}\int_0^tdt'E_n(t')}|n;t\rangle\langle n;0|
%\end{equation}
%provided we were able to solve the eigenvalue problem $H(t)|n;t\rangle=E_n(t)|n;t\rangle$ and the geometric part of the phase is given by $\dot\gamma_n(t)=\langle n;t|i\partial_t|n;t\rangle$.

The main question about this approximation, arisen in these last years, is its condition of validity. Condition (\ref{eq:cond}) seems to arise quite naturally from higher order terms of the adiabatic approximation. Indeed, given the Hamiltonian 
% 16 November 2011 Asked by the referee
\cite{most,fra1}
\begin{equation}
   H'(t)=-\sum_{n,m,n\ne m}e^{i[\gamma_n(t)-\gamma_m(t)]}e^{-\frac{i}{\hbar}\int_0^tdt'[E_m(t')-E_n(t')]}
   \langle m;t|i\hbar\partial_t|n;t\rangle|m;0\rangle\langle n;0|,
\end{equation}
higher order corrections to the unitary evolution of adiabatic approximation are promptly obtained and we have the identity
\begin{equation}
\label{eq:adexp}
   U(t,0)=U_0(t,0){\cal T}\exp\left(-\frac{i}{\hbar}\int_0^tdt'H'(t')\right)
\end{equation}
being ${\cal T}$ the time ordering operator. 
% 16 November 2011 Added due to referee's request
Eq.(\ref{eq:adexp}) is equivalent to
\begin{eqnarray}
   U(t,0)&=&\sum_n e^{i\gamma_n(t)}e^{-\frac{i}{\hbar}\int_0^t dt'E_n(t')}|n;t\rangle\langle n;0| \\ \nonumber
   &+&\frac{i}{\hbar}\sum_{n,k,n\ne k}
   e^{i\gamma_k(t)}e^{-\frac{i}{\hbar}\int_0^t dt'E_k(t')}
   \int_0^t dt'e^{i[\gamma_n(t')-\gamma_k(t')]}e^{-\frac{i}{\hbar}\int_0^{t'}dt''[E_k(t'')-E_n(t'')]}\times \\ \nonumber
   &&\langle k;t'|i\hbar\partial_{t'}|n;t'\rangle|k;t\rangle\langle n;0|+\ldots    
\end{eqnarray} 
%This
that 
has exactly the structure of a Dyson series. So, apparently, condition (\ref{eq:cond}) should grant that 
% 16 November 2011 minor correction
%these 
the
higher order corrections are negligible
% 16 November 2011 added
impeding in this way transitions to other states.

Some years ago, Mostafazadeh \cite{most} proved that the adiabatic series has a parameter that characterizes the expansion. This can be easily understood as Hamiltonians represent physical systems and so have physical parameters. An expansion for the unitary evolution will contain dimensionless ratios of these parameters. Indeed, we proved that the adiabatic expansion is dual to small perturbation Dyson series\cite{fra1}. This can be seen from eq.(\ref{eq:T}). We can write down the following problem for the Schr\"odinger equation
\begin{equation}
\label{eq:lambda}
  \lambda H(t)|\psi\rangle=i\hbar\partial_t|\psi\rangle
\end{equation}
being $\lambda$ an arbitrary dimensionless parameter. In the limit of $\lambda\rightarrow 0$ this can be immediately solved with a Dyson series
\begin{equation}
\label{eq:dyson}
  U(t,0)=I-\frac{i}{\hbar}\lambda\int_0^t dt'H(t')-\frac{1}{\hbar^2}\lambda^2\int_0^tdt'\int_0^{t''}dt'dt''H(t')H(t'')+\ldots
\end{equation}
or, in more compact notation,
\begin{equation}
  U(t,0)={\cal T}\exp\left(-\frac{i}{\hbar}\lambda\int_0^tdt'H(t')\right).
\end{equation}
In the opposite limit $\lambda\rightarrow\infty$, one recover eq.(\ref{eq:T}) from eq.(\ref{eq:lambda}) with the dimensional parameter $T$ interchanged with the dimensionless parameter $\lambda$ but in the same increasingly large limit. So, one has immediately the adiabatic series (\ref{eq:adexp}) by invoking the proof of the adiabatic theorem. These series are dual each other as the expansion parameter of the Dyson series is the inverse of that of the dual series. This appears very clear if we would like to attach a meaning to the dual Dyson series in the limit $\lambda\rightarrow\infty$. This fact has a serious implication for the adiabatic approximation as normally considered in textbooks: Being an ordinary perturbation series this is plagued with secularities. A secularity is a term in a perturbation expansion that increases without bound as the independent variable is increased. These are so dubbed as firstly were recognized in perturbation calculations for celestial mechanics. The very existence of such terms in higher order corrections of the adiabatic approximations implies immediately that the condition (\ref{eq:cond}) cannot grant, as is, the validity of the approximation itself. In fact, it is a normal way out from the failure of this condition to take some bounds in time for the validity of the adiabatic approximation. This is what was normally done when firstly secularities were met in perturbation series but, of course, is not a proper solution. The reason is that one has to understand where the problem comes from and to solve it correctly. This will be shown explicitly in the next section.

The main result we will give here is that the adiabatic theorem represents just the leading order of a perturbation series. This series displays all the common drawbacks of a standard perturbation series making the original condition for the adiabatic theorem to hold, eq.(\ref{eq:cond}), generally useless for the most common applications. Of course, this could be escaped in the original and subsequent formulations of this theorem as a concrete understanding was achieved just in recent years \cite{most,fra1}. Removal of these drawbacks recovers a correct condition but makes this dependent on the particular problem at hand.

\section{\label{sec3}Examples}

\subsection{Validating the theoretical analysis}

In order to show how the above stated theorem on dual series works, we consider a well-known physical model and compare Taylor expansions of its exact solution with the explicit computations obtained from the Dyson series and its dual (the adiabatic expansion). The model we are going to consider is the Jaynes-Cummings model, widely used in quantum optics
% 16 November 2011 This was mostly due 
and we follow strictly the treatment given in \cite{frad}.
% After referee's comment 30 January 2012
It is important to see from this example that resonances are not needed to see the appearance of secular terms in the series, as correctly advocated in Ref. \cite{comp}.
The Hamiltonian is \cite{scul}
\begin{equation}
    H_{JC}=\omega a^\dagger a
    +\frac{\omega_0}{2}(|2\rangle\langle 2|-|1\rangle\langle 1|)
    +g(|2\rangle\langle 1|a^\dagger +|1\rangle\langle 2|a).
\end{equation}
This represents a two-level atom coupled with a single mode radiation of frequency $\omega$ and coupling $g$.

One can rewrite the Hamiltonian in the interaction picture obtaining
\begin{equation}
    H_{JC}^{(I)}=g(e^{i\Delta t}|2\rangle\langle 1|a^\dagger +
    e^{-i\Delta t}|1\rangle\langle 2|a)
\end{equation}
being $\Delta=\omega_0-\omega$ the detuning that here we take different from $0$ for the sake of generality. It is immediate to realize that the physical parameter for this model is the ratio $\frac{g}{\Delta}$. This means that any perturbation series and its dual will have this parameter or its inverse as a development parameter. The exact solution of the Schr\"{o}dinger equation in interaction picture can be found by looking for a solution in the form
\begin{equation}
    |\psi\rangle_I=\sum_n c_{1,n+1}(t)|1,n+1\rangle+c_{2,n}(t)|2,n\rangle
\end{equation}
being $n$ the photon number. The probability amplitudes are given by \cite{scul}
\begin{eqnarray}
    c_{1,n+1}(t)&=&\left\{c_{1,n+1}(0)\left[
    \cos\left(\frac{\Omega_nt}{2}\right)+\frac{i\Delta}{\Omega_n}
    \sin\left(\frac{\Omega_nt}{2}\right)\right]-\frac{2ig\sqrt{n+1}}
    {\Omega_n}c_{2,n}(0)\sin\left(\frac{\Omega_nt}{2}\right)
    \right\}e^{-i\Delta t/2} \nonumber \\
    & & \label{eq:sjc} \\
    c_{2,n}(t)&=&\left\{c_{2,n}(0)\left[
    \cos\left(\frac{\Omega_nt}{2}\right)-\frac{i\Delta}{\Omega_n}
    \sin\left(\frac{\Omega_nt}{2}\right)\right]-\frac{2ig\sqrt{n+1}}
    {\Omega_n}c_{1,n+1}(0)\sin\left(\frac{\Omega_nt}{2}\right)
    \right\}e^{i\Delta t/2} \nonumber
\end{eqnarray}
being $\Omega_n=\sqrt{\Delta^2+{\cal R}^2_n}$ and ${\cal R}_n=2g\sqrt{n+1}$ the Rabi frequency. The Dyson series is obtained by expanding the above solution in Taylor series of $\lambda=\frac{{\cal R}_n}{\Delta}$ giving till second order
\begin{eqnarray}
    c_{1,n+1}(t)&=&\left\{c_{1,n+1}(0)\left[
    1+
    i\frac{\lambda^2}{4}
    \left(\Delta t +
    i(1-e^{-i\Delta t})\right)
    \right]
    -\frac{\lambda}
    {2}c_{2,n}(0)(1-e^{-i\Delta t})+O(\lambda^3)
    \right\} \nonumber \\
    & & \label{eq:sjca} \\
    c_{2,n}(t)&=&\left\{c_{2,n}(0)\left[
    1-i\frac{\lambda^2}{4}
    \left(\Delta t +i
    (e^{i\Delta t}-1)\right)
    \right]
    -\frac{\lambda}
    {2}c_{1,n+1}(0)(e^{i\Delta t}-1)+O(\lambda^3)
    \right\}. \nonumber
\end{eqnarray}
The secularity in this perturbation series can be easily recognized as this is the term that grows without bound in the
limit $t\rightarrow\infty$. In perturbation theory, unless we are not able to get rid of the secularity, such a series is not very useful and does not say too much about the true solution. Several methods exist to resum secularities as e.g. the renormalization group methods described in Ref.\cite{fra2}. But here, the problem can be easily traced back to the Taylor expansion of the functions $\sin(\sqrt{1+\epsilon^2}t)$ in $\epsilon$, having $\sqrt{1+\epsilon^2}=1+\frac{\epsilon^2}{2}+O(\epsilon^4)$. This can be eliminated by simply substituting $\Delta$ with $\Delta + \frac{{\cal R}_n^2}{2\Delta}$ everywhere in the approximate solution into the exponentials of eq.(\ref{eq:sjca}). The expansion (\ref{eq:sjca}) can be immediately obtained through the Dyson series (\ref{eq:dyson}). So, as expected, this series gives an analysis of the Jaynes-Cummings model when the detuning $\Delta$ is enough larger than the Rabi frequency ${\cal R}_n$. We have easily disposed of the secular term that made the series useless.

Now, let us repeat the above discussion in the opposite limit with the Rabi frequency larger than the detuning. Again, by Taylor expanding the exact solution one has
\begin{eqnarray}
    c_{1,n+1}(t)&=&\left\{c_{1,n+1}(0)\left[
    \cos\left(\frac{{\cal R}_n}{2}t\right)
    +\frac{i}{\lambda}
    \sin\left(\frac{{\cal R}_n}{2}t\right)-
    \frac{1}{2\lambda^2}\frac{{\cal R}_n}{2}t
    \sin\left(\frac{{\cal R}_n}{2}t\right)
    \right]\right. \nonumber \\
    &-&
    \left.ic_{2,n}(0)
    \left[\sin\left(\frac{{\cal R}_n}{2}t\right)
    -\frac{1}{2\lambda^2}
    \left(\sin\left(\frac{{\cal R}_n}{2}t\right)-\frac{{\cal R}_n}{2}t
    \cos\left(\frac{{\cal R}_n}{2}t\right)\right)\right] +
    O\left(\frac{1}{\lambda^3}\right)
    \right\}e^{-i\Delta t/2} \nonumber \\
    & & \label{eq:sjcb} \\
    c_{2,n}(t)&=&\left\{c_{2,n}(0)\left[
    \cos\left(\frac{{\cal R}_n}{2}t\right)-\frac{i}{\lambda}
    \sin\left(\frac{{\cal R}_n}{2}t\right)
    -\frac{1}{2\lambda^2}\frac{{\cal R}_n}{2}t
    \sin\left(\frac{{\cal R}_n}{2}t\right)
    \right]\right. \nonumber \\
    &-& \left.ic_{1,n+1}(0)
    \left[\sin\left(\frac{{\cal R}_n}{2}t\right)
    -\frac{1}{2\lambda^2}
    \left(\sin\left(\frac{{\cal R}_n}{2}t\right)-\frac{{\cal R}_n}{2}t
    \cos\left(\frac{{\cal R}_n}{2}t\right)\right)\right]+
    O\left(\frac{1}{\lambda^3}\right)
    \right\}
    e^{i\Delta t/2} \nonumber
\end{eqnarray}
with the same problem of a secularity at second order. This series can be immediately obtained by the adiabatic expansion (dual Dyson series) showing what could seem an unexpected result from the adiabatic approximation, but in complete agreement with the results of Ref.\cite{fra1} and our preceding discussion. To compute the dual Dyson series we need the eigenstates and eigenvalues of $H_{JC}^{(I)}$. It is easily found that for the eigenvalue $g\sqrt{n+1}$ we have the eigenstate
\begin{equation}
    |a,n,t\rangle=\frac{1}{\sqrt{2}}
    (e^{-i\Delta t}|1,n+1\rangle+|2,n\rangle)
\end{equation}
and for the eigenvalue $-g\sqrt{n+1}$ we have the eigenstate
\begin{equation}
    |b,n,t\rangle=\frac{1}{\sqrt{2}}
    (|1,n+1\rangle-e^{i\Delta t}|2,n\rangle).
\end{equation}
In quantum optics, these are normally dubbed ``dressed states'' \cite{scul}. Geometric phases are then easily computed to give
\begin{eqnarray}
    \dot{\gamma}_a &=&
    \langle a,n,t|i\frac{\partial}{\partial t}|a,n,t\rangle =
    \frac{\Delta}{2} \nonumber \\
    & & \\
    \dot{\gamma}_b &=&
    \langle b,n,t|i\frac{\partial}{\partial t}|b,n,t\rangle =
    -\frac{\Delta}{2}.    \nonumber \\
\end{eqnarray}
Then, after some algebra using the dressed states computed above, the unitary evolution operator is given by,
\begin{eqnarray}
    U_0(t)&=&
    e^{i\frac{\Delta}{2}t-ig\sqrt{n+1}t}|a,n,t\rangle\langle a,n,0|+
    e^{-i\frac{\Delta}{2}t+ig\sqrt{n+1}t}|b,n,t\rangle\langle b,n,0|
    \nonumber \\
        &=&\cos\left(\frac{{\cal R}_n}{2}t\right)
    (e^{-i\frac{\Delta}{2}t}|1,n+1\rangle\langle 1,n+1| +
     e^{i\frac{\Delta}{2}t}|2,n\rangle\langle 2,n|) \label{eq:u0} \\
     &-&i\sin\left(\frac{{\cal R}_n}{2}t\right)
     (e^{-i\frac{\Delta}{2}t}|1,n+1\rangle\langle 2,n| +
      e^{i\frac{\Delta}{2}t}|2,n\rangle\langle 1,n+1|) \nonumber
\end{eqnarray}
that, for $|\psi(0)\rangle=c_{1,n+1}(0)|1,n+1\rangle+c_{2,n}(0)|2,n\rangle$, gives
\begin{eqnarray}
    |\psi(t)\rangle_I &\approx&
    \left[\cos\left(\frac{{\cal R}_n}{2}t\right)c_{1,n+1}(0)
     -i\sin\left(\frac{{\cal R}_n}{2}t\right)c_{2,n}(0)\right]
     e^{-i\frac{\Delta}{2}t}|1,n+1\rangle \nonumber \\
    & & \\
    &+&\left[\cos\left(\frac{{\cal R}_n}{2}t\right)c_{2,n}(0)
     -i\sin\left(\frac{{\cal R}_n}{2}t\right)c_{1,n+1}(0)\right]
     e^{i\frac{\Delta}{2}t}|2,n\rangle \nonumber
\end{eqnarray}
that is the exact form of eqs.(\ref{eq:sjcb}) when higher order terms beyond the leading one are neglected, i.e. when $\lambda\rightarrow\infty$, as expected from the results of Ref.\cite{fra1} and our discussion.

We improve our computation going to higher orders. Again, using the above expressions for the dressed states one gets
\begin{eqnarray}
    H'(t)&=&-\frac{\Delta}{2}
    \left[\cos\left({\cal R}_n t\right)
    (|1,n+1\rangle\langle 1,n+1| -
     |2,n\rangle\langle 2,n|)\right. \nonumber \\
     & & \\
     &-&\left.i\sin\left({\cal R}_n t\right)
     (|1,n+1\rangle\langle 2,n| -
      |2,n\rangle\langle 1,n+1|)\right] \nonumber
\end{eqnarray}
so that, the first order correction to the leading order evolution operator $U_0(t)$ of eq.(\ref{eq:u0}) is given by
\begin{eqnarray}
    U_1(t)=-iU_0(t)\int_0^t dt_1H'(t_1)=
    i\frac{1}{\lambda}\sin\left(\frac{{\cal R}_n}{2}t\right)
    (e^{-i\frac{\Delta}{2}t}|1,n+1\rangle\langle 1,n+1|-
     e^{i\frac{\Delta}{2}t}|2,n\rangle\langle 2,n|)
\end{eqnarray}
that gives the first order correction
\begin{eqnarray}
    |\delta_1\psi(t)\rangle_I=
    i\frac{1}{\lambda}\sin\left(\frac{{\cal R}_n}{2}t\right)
    (e^{-i\frac{\Delta}{2}t}c_{1,n+1}(0)|1,n+1\rangle-
     e^{i\frac{\Delta}{2}t}c_{2,n}(0)|2,n\rangle)
\end{eqnarray}
again in agreement with the Taylor expansion as given in eqs.(\ref{eq:sjcb}), to order $\frac{1}{\lambda}$. So, in the same way we have at the second order
\begin{eqnarray}
\label{eq:U2}
    U_2(t)&=&-U_0(t)\int_0^t dt_1 H'(t_1)
    \int_0^{t_1}dt_2 H'(t_2)= \nonumber \\
    & &i\frac{1}{2\lambda^2}
    \left\{\left[\sin\left(\frac{{\cal R}_n}{2}t\right)
    -\frac{{\cal R}_n}{2}t
    \cos\left(\frac{{\cal R}_n}{2}t\right)\right]
    (e^{i\frac{\Delta}{2}t}|2,n \rangle\langle 1,n+1|+
    e^{-i\frac{\Delta}{2}t}|1,n+1 \rangle\langle 2,n|)\right. \\
    &-&\left.i\frac{{\cal R}_n}{2}t
    \sin\left(\frac{{\cal R}_n}{2}t\right)
    (e^{-i\frac{\Delta}{2}t}|1,n+1\rangle\langle 1,n+1|+
     e^{i\frac{\Delta}{2}t}|2,n\rangle\langle 2,n|)\right\} \nonumber
\end{eqnarray}
then, one has
\begin{eqnarray}
    |\delta_2\psi(t)\rangle_I&=&
    i\frac{1}{2\lambda^2}
    \left\{\left[\sin\left(\frac{{\cal R}_n}{2}t\right)
    -\frac{{\cal R}_n}{2}t
    \cos\left(\frac{{\cal R}_n}{2}t\right)\right]
    (e^{i\frac{\Delta}{2}t}c_{1,n+1}(0)|2,n\rangle+
     e^{-i\frac{\Delta}{2}t}c_{2,n}(0)|1,n+1\rangle)\right.
     \nonumber \\
     & & \\
    &-&i\left.\frac{{\cal R}_n}{2}t
    \sin\left(\frac{{\cal R}_n}{2}t\right)
    (e^{-i\frac{\Delta}{2}t}c_{1,n+1}(0)|1,n+1\rangle+
     e^{i\frac{\Delta}{2}t}c_{2,n}(0)|2,n\rangle)\right\}. \nonumber
\end{eqnarray}
The agreement with the Taylor expansion as given in eqs.(\ref{eq:sjcb}), to order $\frac{1}{\lambda^2}$, is complete. This series is dual to the Dyson series as the development parameter is $1/\lambda$. It is clear from these computations, by inspecting eq.(\ref{eq:U2}), that the adiabatic expansion produces secular terms and these terms impede a correct evaluation of the validity bounds through eq.(\ref{eq:cond}). It is also clear that imposing bounds on the range of the time variable cannot solve the question but, rather, secular terms must be resummed to make the series meaningful and only after this, one should establish the range of validity of the adiabatic approximation.

\subsection{\label{sec3b} More common examples}

In order to give one of these examples of failure, we consider the one discussed in \cite{tong,comp}. This example has the property to be exactly solvable and the adiabatic approximation can be immediately compared with the exact solution. The Hamiltonian, firstly due to Schwinger \cite{schw}, is
\begin{equation}
   H=-\frac{\hbar\omega_0}{2}(\sigma_x\sin\theta\cos\omega t+\sigma_y\sin\theta\sin\omega t+\sigma_z\cos\theta).
\end{equation}
Instantaneous eigenvectors and eigenvalues are immediately obtained as
\begin{equation}
 E_{0,1} = \mp \frac{\hbar\omega_0}{2}, \qquad
 |0;t\rangle {=} \left( \begin{array}{c} e^{-i\frac{\omega}{2}t}\cos\frac{\theta}{2} \\
    e^{i\frac{\omega}{2}t}\sin\frac{\theta}{2}
 \end{array}\right), \qquad
 |1;t\rangle {=}\left( \begin{array}{c} e^{-i\frac{\omega}{2}t}\sin\frac{\theta}{2} \\ -e^{i\frac{\omega}{2}t}\cos\frac{\theta}{2}
 \end{array} \right),
\end{equation}
and the geometrical phases are given by $\gamma_{0,1}=\pm\frac{\omega}{2}\cos\theta$. So, the adiabatic approximation provides the unitary evolution
\begin{equation}
   U_0(t,0)=e^{i\frac{\omega}{2}\cos\theta t}e^{i\frac{\omega_0}{2}t}|0;t\rangle\langle 0;0|
   +e^{-i\frac{\omega}{2}\cos\theta t}e^{-i\frac{\omega_0}{2}t}|1;t\rangle\langle 1;0|
\end{equation}
that is
\begin{widetext}
\begin{eqnarray}
U_0(t,0)=\left(\begin{array}{cc}
\left(\cos\frac{1}{2}\tilde\omega t+i\cos\theta\sin\frac{1}{2}\tilde\omega t\right)e^{-i\frac{\omega t}{2}} & 
i\sin\theta\sin\frac{\tilde\omega t}{2}e^{-i\frac{\omega t}{2}}\\
i\sin\theta\sin\frac{\tilde\omega t}{2}e^{i\frac{\omega t}{2}}
&\left(\cos\frac{1}{2}\tilde\omega t-i\cos\theta\sin\frac{1}{2}\tilde\omega t\right)e^{i\frac{\omega t}{2}} \end{array}\right)
\label{eq:u0tong}
\end{eqnarray}
\end{widetext}
being $\tilde\omega=\omega_0+\omega\cos\theta$.
The exact solution takes the form \cite{tong}
\begin{widetext}
\begin{eqnarray}
U(t,0)=\left(\begin{array}{cc}
(\cos\frac{\overline\omega t}{2}+i\frac{\omega+\omega_0\cos\theta}{\overline\omega}\sin\frac{\overline\omega t}{2})e^{-i\frac{\omega t}{2}} & i\frac{\omega_0\sin\theta}{\overline\omega}\sin\frac{\overline\omega t}{2}e^{-i\frac{\omega t}{2}}\\
i\frac{\omega_0\sin\theta}{\overline\omega}\sin\frac{\overline\omega t}{2}e^{i\frac{\omega t}{2}}
&(\cos\frac{\overline\omega t}{2}-i\frac{\omega+\omega_0\cos\theta}{\overline\omega}\sin\frac{\overline\omega t}{2})e^{i\frac{\omega t}{2}} \end{array}\right),
\label{u}
\end{eqnarray}
\end{widetext}
being $\overline\omega=\sqrt{\omega_0^2+\omega^2+2\omega_0\omega\cos\theta}$. Now, we can recover eq.(\ref{eq:u0tong}) from the exact solution quite easily with the condition $\omega_0\gg\omega$. In this case, $\overline\omega\approx\tilde\omega$ and it is consistent with the condition for adiabatic approximation that is easily obtained from (\ref{eq:cond}) to be $\omega_0\gg\omega\sin\theta$. So, we realize that the adiabatic approximation is just a standard series expansion as any other perturbation expansion but we discuss this below. Here, we just note that adiabaticity condition proves to be not sufficient to grant the applicability of this approximation. This can be immediately realized if we try to apply it on the exact unitary evolution to get higher order corrections. These cannot be neglected if we take the extreme of the interval of time evolution of the adiabatic approximation $T$ large enough and so our condition fails to grant applicability. 
%This point is essential to understand the ongoing discussion about this matter in the literature and our following analysis.
Now, using eigenvalues, eigenvectors and the geometric phases already computed, we are able to get the Hamiltonian for higher order corrections
\begin{equation}
   H'=-\frac{1}{2}\hbar\omega\sin\theta e^{-i(\omega_0+\omega\cos\theta)t}|0;0\rangle\langle 1;0|+c.c.
\end{equation}
We see that when the resonance condition $\omega_0+\omega\cos\theta\approx 0$ is met, a secularity is immediately produced causing the failure of the standard adiabaticity condition. Due to the presence of these terms, the adiabaticity condition cannot be neither sufficient nor necessary unless these terms are properly summed away. Particularly, for this example, even if the resonance condition is not met, going to higher orders will give rise to powers of time variable making the adiabatic series more and more unreliable. At resonance one gets
\begin{equation}
 H' = -\frac{\hbar\omega}{2}\sin^2\theta\sigma_z+\frac{\hbar\omega}{2}\sin\theta\cos\theta\sigma_x
\end{equation}
and the secular contribution is then
\begin{equation}
   -i\int_0^tdt'H'(t')=i\frac{\hbar\omega}{2}\sin\theta(\sin\theta\sigma_z-\cos\theta\sigma_x) t.
\end{equation}
The existence of such a term makes useless the adiabaticity condition.  But when we are off the resonance, the situation is the same as the condition for adiabatic evolution just fails. In fact, the second order term for the adiabatic evolution can be easily evaluated as
\begin{eqnarray}
   -\int_0^tdt'H'(t')\int_0^{t''}dt''H'(t'')&=&-\frac{1}{4}\hbar^2\omega^2\sin^2\theta\int_0^tdt'\int_0^{t''}dt''
   \left[e^{-i(\omega_0+\omega\cos\theta)(t'-t'')}|0;0\rangle\langle 0;0|\right. \\ \nonumber
   &&+\left.e^{i(\omega_0+\omega\cos\theta)(t'-t'')}|1;0\rangle\langle 1;0|\right] \\ \nonumber
   &&=\frac{1}{4}\hbar^2\omega^2\sin^2\theta \frac{t}{\omega_0+\omega\cos\theta}.\left[|0;0\rangle\langle 0;0|
   -|1;0\rangle\langle 1;0|\right]+n.s.t.
\end{eqnarray}
with n.s.t. meaning ``not secular terms''. So, for this kind of model, the next-next-to-leading order produces a secularity in any case that makes the adiabatic series fail. This term must be resummed to get a fine adiabatic expansion. We just note that we can read off the adiabatic condition in the form $|\omega\sin\theta/(\omega_0+\omega\cos\theta)|\approx|\omega\sin\theta/\overline\omega|\ll 1$, in agreement with the conclusions given in \cite{comp}. This is also the expansion parameter of the series. When $t$ is increased enough, the series fails and the adiabaticity condition becomes useless.

%In order to understand why the normal condition for adiabaticity fails, we consider a couple of examples taken from the literature. The simplest
Another simple and largely known example is the one given in \cite{amin}. The Hamiltonian is
\begin{equation}
    H=-\frac{\epsilon}{2}\sigma_z-V\cos\omega_0t\sigma_x.
\end{equation}
Here and in the following $\sigma_x,\ \sigma_y,\ \sigma_z$ are the Pauli matrices.
This model is widely studied and it is known that it undergoes suppression of tunneling at the leading order but the corresponding condition is not exact being lost by higher order corrections. All this is computed through the adiabatic series exposed in sec.\ref{sec2} \cite{fra3,fra4} but see also \cite{bara} for a different approach. The essence of the approach to recover the adiabatic approximation, as already stated above, is the elimination of the secularities into the adiabatic perturbation series. In ref.\cite{amin} it is claimed that rotating wave approximation should apply but this is not consistent with the adiabatic approximation, as we are going to show in a moment. In fact, it is not convenient to apply directly the adiabatic approximation to this Hamiltonian. Rather, we do a change of representation going to the interaction picture. This will give
\begin{equation}
    H_I=-V\cos\omega_0t\sigma_xe^{i\frac{\epsilon}{\hbar}\sigma_zt}
\end{equation}
and so we get the eigenstates and eigenvalues
\begin{equation}
 E_{0,1} = \pm V\cos\omega_0 t, \qquad
 |0;t\rangle_I {=} \frac{1}{\sqrt{2}}\left( \begin{array}{c} 1 \\
 -e^{\frac{i}{\hbar}\epsilon t}
 \end{array}\right), \qquad
 |1;t\rangle_I {=} \frac{1}{\sqrt{2}}\left( \begin{array}{c} e^{-\frac{i}{\hbar}\epsilon t} \\ 1
 \end{array} \right),
\end{equation}
and for the geometric phases we get $\dot\gamma_{0,1}=\mp\frac{\epsilon}{2}$. So, in the interaction picture, the unitary adiabatic evolution is given by
\begin{equation}
   U_{0I}(t,0)=e^{-\frac{i}{2\hbar}\epsilon t}e^{-\frac{i}{\hbar}\frac{V}{\omega_0}\sin\omega_0 t}|0;t\rangle_I\ _I\langle 0;0|+
   e^{\frac{i}{2\hbar}\epsilon t}e^{\frac{i}{\hbar}\frac{V}{\omega_0}\sin\omega_0 t}|1;t\rangle_I\ _I\langle 1;0|
\end{equation}
and finally
\begin{equation}
   U_0(t,0)=e^{i\frac{\epsilon}{2\hbar}\sigma_zt}U_{0I}(t,0)
\end{equation}
that gives
\begin{equation}
   U_0(t,0)=e^{-\frac{i}{\hbar}\frac{V}{\omega_0}\sin\omega_0 t}|0;t\rangle\langle 0;0|+
   e^{\frac{i}{\hbar}\frac{V}{\omega_0}\sin\omega_0 t}|1;t\rangle\langle 1;0|
\end{equation}
being
\begin{equation}
 |0;t\rangle {=} \frac{1}{\sqrt{2}}\left( \begin{array}{c} 1 \\
 -1
 \end{array}\right), \qquad
 |1;t\rangle {=} \frac{1}{\sqrt{2}}\left( \begin{array}{c} 1 \\ 1
 \end{array} \right),
\end{equation}
that are independent on time and are eigenstates of $\sigma_x$. In this way we can immediately compute the Hamiltonian for higher corrections obtaining
\begin{equation}
   H'=-\frac{\epsilon}{2}\sigma_ze^{\frac{i}{\hbar}\frac{V}{\omega_0}\sin\omega_0 t\sigma_x}.
\end{equation}
With this Hamiltonian we can evaluate the correction to be neglected for the adiabatic approximation to hold. Indeed, using the identity
\begin{equation}
   e^{\frac{i}{\hbar}\frac{V}{\omega_0}\sin\omega_0 t\sigma_x}=
   \sum_{n=0}^\infty J_n\left(\frac{V}{\hbar\omega_0}\right)e^{in\omega_0t\sigma_x}
\end{equation}
being $J_n$ the Bessel functions of integer order, we recognize that
\begin{equation}
   H'=-\frac{\epsilon}{2}J_0\left(\frac{V}{\hbar\omega_0}\right)\sigma_z+\ldots
\end{equation}
and this constant term will give rise to a secularity producing a failure into the adiabaticity condition. But this term can be just resummed away to give a proper unitary evolution \cite{fra3,fra4}
\begin{equation}
   U(t,0)\approx e^{\frac{i}{\hbar}\frac{V}{\omega_0}\sin\omega_0 t\sigma_x}
   e^{\frac{i}{\hbar}\frac{\epsilon}{2}J_0\left(\frac{V}{\hbar\omega_0}\right)t\sigma_z}
\end{equation}
and the approximation holds provided $\epsilon J_n(V/\hbar\omega_0)/\hbar\omega_0\ll 1$ for any $n>0$. This is again a strong coupling expansion as this condition is granted, for any $n$, provided $V\gg\hbar\omega_0$ also at resonance and this rules out rotating wave approximation. But we turn our attention to the adiabatic theorem, we now realize immediately that, after resummation of the secularities, this holds unconditionally while the usual adiabaticity condition, $V\omega_0|\sin\omega_0 t|/\epsilon^2\ll 1$ (see \cite{amin}), just fails.  So, to verify how and when the adiabatic theorem holds we need to go to the next-to-leading order, at least, and resum secular terms.

%The next example we want to consider is the Schwinger model considered in \cite{tong,comp} already discussed in sec.\ref{sec1}. 

%So, following \cite{amin}, one has
%\begin{equation}
% E_{0,1} = \mp {1\over 2} \Omega, \qquad
% |E_{0,1};t\rangle {=} \left( \begin{array}{c} \alpha^\pm \\
% \pm\alpha^\mp
% \end{array}\right), \quad
% |E_1\rangle {=} \left( \begin{array}{c} \alpha^- \\ -\alpha^+
% \end{array} \right),
%\end{equation}
%being $\Omega = \sqrt{\epsilon^2 + 4V^2 \sin^2 \omega_0 t}$ and $\alpha^\pm = \sqrt{(\Omega \pm \epsilon)/2 \Omega}$. This says us that the adiabatic evolution at the leading order has the unitary operator
%\begin{equation}
%   U_0(t,0)=\exp\left(\frac{i}{2\hbar}\int_0^tdt'\Omega(t')\right)|E_0;t\rangle\langle E_0;0|
%   +\exp\left(-\frac{i}{2\hbar}\int_0^tdt'\Omega(t')\right)|E_1;t\rangle\langle E_1;0|
%\end{equation}
%being $\sigma_z|E_{0,1};0\rangle=\mp|E_{0,1};0\rangle$. Now, from these we get the Hamiltonian to get higher order corrections to the adiabatic series as
%\begin{equation}
%   H'=-\frac{1}{2}\frac{\omega\epsilon\cot\omega t}{(\epsilon^2+4V^2\sin^2\omega t)}
%   e^{\frac{2i}{\hbar}\int_0^t\sqrt{\epsilon^2+4V^2\sin^2\omega t}}|E_0;0\rangle\langle E_1;0|+c.c.
%\end{equation}

\section{\label{sec4}Conclusions}

We have shown the origin of the difficulties with the standard validity condition of the adiabatic approximation. This just arises from the perturbative nature of this approximation that makes enter into play some singular behaviors of the expansion. Disposing of these terms puts out the right conditions for the validity of the adiabatic theorem. Besides, a clear understanding of the expansion parameter of the series should be achieved to state properly a validity condition.

As a side result, we are able to produce a perturbation series for a strongly perturbed Schr\"odinger equation. This is a quite general approach to manage equations with a term going formally to infinity.

Our hope is that this should give rise to a new understanding of this old approximation providing some new insights in fields of physics where applications are at hand.

\begin{acknowledgments}
I would like to thank Kazuyuki Fujii for pointing me out some corrections to the equations of sec.\ref{sec3b}.
\end{acknowledgments}

\end{document}